\documentclass{IEEEcsmag}

\usepackage[breaklinks,colorlinks,urlcolor=blue,linkcolor=blue,citecolor=blue]{hyperref}%
\usepackage{xurl}

\usepackage{upmath}

\usepackage{wrapfig}
\usepackage{xcolor}
\usepackage{tcolorbox}

\newlength\mylen
\newcommand{\sidebox}[5]{{\color{#3}\noindent\colorbox{#2}{\parbox{#1}{
\begin{center}\textbf{#4} \end{center}

#5
}}}}

\jname{IEEE Computing in Science \& Engineering}
\pubyear{2022}

\begin{document}


\title{Structured and Unstructured Teams for Research Software Development at the Netherlands eScience Center}

\author{Carlos Martinez-Ortiz, Rena Bakhshi, Yifat Dzigan, Nicolas Renaud, Faruk Diblen, Berend Weel, Maarten van Meersbergen, Niels Drost, Sven van der Burg, and Fakhereh (Sarah) Alidoost}
\affil{Netherlands eScience Center, Amsterdam, The Netherlands}

\markboth{C. Martinez-Ortiz \emph{et al.}}{Structured and Unstructured Teams...}

\begin{abstract}
This paper presents the types of teams that are currently in place at the Netherlands eScience Center. We categorize our teams into four different types: Project Teams, Collectives, Agile Teams and Scrum Teams. We provide a brief description of each team type and share stories from teams themselves to reflect on the strengths and shortcomings of each model. From our observation, we conclude that the type of team that is most suitable for each project depends on the specific needs of that project. More importantly, different types of teams are suitable for the different types of people working at the Netherlands eScience Center.\\[2ex]

\\[1ex]

This is a preprint of the accepted article \url{https://doi.org/10.1109/MCSE.2022.3167448}.
\end{abstract}

\maketitle

\chapterinitial{Research} software projects have grown in complexity and size, and often require teams of people rather than developer working alone. Knowledge transfer and robustness are other reasons to have two or more people on a project. The question of how to structure and manage such teams has received little attention from scientific community until now, but that is changing. At the Netherlands eScience Center, we have made deliberate efforts in this topic.

The Netherlands eScience Center\footnote{\url{https://www.esciencecenter.nl/}} is the Dutch national center of expertise for research software engineering. Together with our academic partners, we develop open-source software and apply these tools to concrete research questions. Our projects cover the entire research landscape ranging from the complex data mining of historical events~\cite{nl-ssh} to large scale computation for climate science~\cite{nl-es}. These projects can also drastically vary in size with some large-scale projects requiring the combined effort of multiple research software engineers (RSEs)~\cite{ewatercycle} to small consultancy projects where RSEs guide external team in their research and development activities\footnote{\url{https://www.esciencecenter.nl/news/the-magic-of-machine-learning-call-winners-announced/}}. Alongside support staff and management, the Center has a pool of about fifty RSEs who work on about fifty running projects in partnership with research groups across the Netherlands and beyond. To cope with the growth of the Center and the complexity of its ambitious mission, RSEs started teaming up to facilitate and organize the execution of our projects. This paper presents the types of teams that are currently in place at the Netherlands eScience Center. Each team type is illustrated by a story where a team reflects on the strengths and shortcomings of their own model of working. We finally offer our own conclusions regarding what type of team suits best the different types of projects and more importantly the people working at the Netherlands eScience Center.

\section{eScience Team Taxonomy}
The notion of a team at the eScience Center started with ‘project teams’ of two RSEs. The upper management wanted to have robustness in cases of personnel change on projects. Over time, the necessity for RSEs to share their time between multiple projects led to several RSEs teaming up to work on multiple projects simultaneously together. The catalyst was a single large project where an ad-hoc team of engineers working on the project part-time was insufficient.

Thus, teams have emerged at the eScience Center in a bottom-up way; each team has defined its own approach to distribute the work among the team members, support each other and collectively advance research through software development. This has allowed the Center to experiment with different formats of research software teams, ranging from a single RSE assigned to a given project, to large, structured groups of RSEs working on several projects. Regardless of the format, all teams interact with external stakeholders, such as group leaders and PhD students, who can also take part in the software development efforts.

\begin{figure*}
\centerline{\includegraphics[width=35pc]{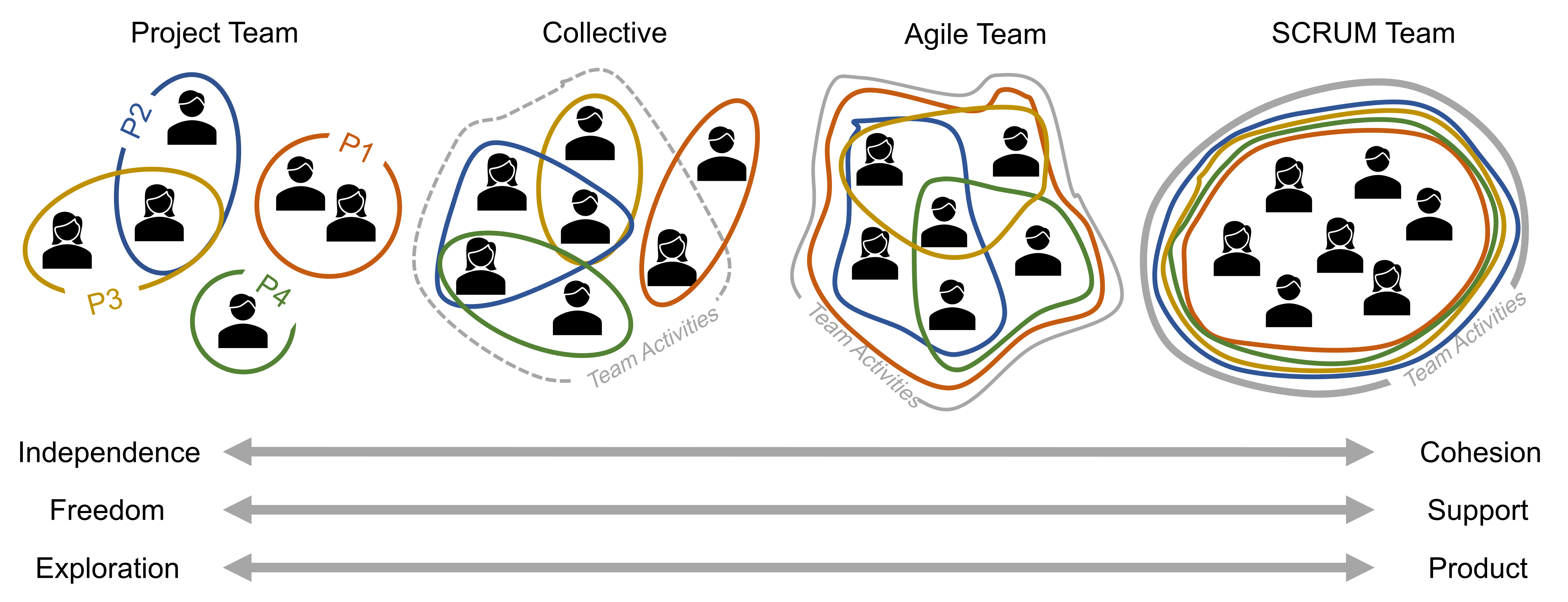}}
\caption{Overview of the diverse types of teams used at the eScience Center ranging from unstructured project teams to structured Scrum teams. All types have pros and cons and suit diverse types of projects and people better.}
\label{fig1}
\end{figure*}

We present an overview of the types of teams that we are experimenting with. We avoid giving a strict definition of what constitutes a team and focus instead on the various ways of working that may suit the needs of different projects and people. Through extensive discussions with team members and managers we have identified four types of teams: Project Teams, Collectives, Agile Teams and Scrum Teams. We use this classification as an abstraction for the purpose of discussion, but do not claim it to be exhaustive. In fact, not all of the current teams at the eScience Center fit exactly in this classification, many teams being somewhere in between two types.  It is also possible for any team to change its modus operandi either permanently or temporarily: for example, several Project Teams teaming up as a Scrum team for the execution of a single sprint, or the members of an Agile team deciding to work independently of each other on personal projects for a week. We briefly present below these four types of teams, also illustrated in Figure~\ref{fig1}. Later we provide short stories for each type to make those definitions more concrete.

\subsection{Project Teams}
Project Teams consist of one or two engineers working on a specific project. The sole purpose of the team is the realization of that project, and the team naturally dissolves itself when the project ends. The realization of the project is usually just one of the focus points for the engineers, who split their time between several projects and therefore several teams. The team members meet when needed to update each other and plan the development of the project. Sprint and pair programming sessions happen sporadically, with team members usually preferring to work asynchronously on the project to avoid scheduling issues.

\subsection{Collectives}
Collectives consist of a group of loosely connected engineers working on a set of similar projects. The similarity between the projects and the frequent interactions between all team members facilitates cooperation. Frequent interactions take place through morning stand ups. Cooperation also takes place through design sessions, pair programming sessions and code reviews. The execution of a given project is, however, done by one or two engineers working on it with minimal contributions from the rest of the team. The loose connection between the team members allows for non-team members to contribute to projects that are executed in the team. Team-wide activities, such as one-week sprint or learning days are organized when opportune.

\subsection{Agile Teams}
A multitude of teams have adopted the Agile philosophy to organize their work and improve team dynamics. These teams vary in sizes ranging from four to eight members. Most teams prefer members committing most of their time to work within the team, although there is flexibility, depending on the role and other responsibilities team members have. Each team works on several projects that generally have some overlap in terms of domain, technology, or both. Daily stand-ups are used to plan the work and keep the entire team updated about the progress of the projects. While collaboration is greatly reinforced by the tight team dynamic, only some team members work on the same project at the same time. Some Agile teams are working in time-constrained sprints of generally two weeks while others prefer a looser format, for example based on the use of a team-wide Kanban board. Note that the Agile manifesto~\cite{agile-manifesto} simply lists several principles rather than prescribed structure or roles for a team.

\subsection{Scrum Teams}
A few teams have adopted (to varying degrees) the Scrum framework to organize their work. Following this approach, Scrum teams work in two- or three-week sprints during which all team members work on a single project. The sprints start with an extensive planning session and end with a sprint review and a retrospective. Daily stand ups are used to continuously update and fine tune the execution of the project. The release of a new feature or product is usually made at the end of each sprint. Team members have defined roles, such as a product owner and a Scrum master, providing clarity with the task distribution and the responsibilities within the team.

\section{eScience Team stories}
In this section we provide short stories from various eScience Center Teams, as examples of the Taxonomy introduced in the previous section. These stories are told by RSEs. They reflect snapshots in time and may be outdated. However, they serve as an illustration of different types of teams and provide valuable insight into our teams’ experiences.

\subsection{Project Teams}
\subsubsection{Project Team 1:} ``My colleague and I teamed up to work on a project where we consult a group of political scientists on a machine learning project for half a year. I commit most of my time to working in another team, but I usually set aside a few hours per week to work on this project. We meet every two weeks with the project partners to discuss progress and new insights. Every week the two of us have a more technical meeting where we update each other on the progress we made individually, unblock each other if we are stuck on something, and together decide on the direction of the project.'' Portfolio discipline: Social sciences.

\subsubsection{Project Team 2:} ``Our team has been just recently formed and is still at the brainstorm stage about a way of working. Unlike any other project team at the Center, our core team works only on one product. The team consists of a front-end developer (part time, externally employed), a back-end developer (full time), a UX developer (only part time in the team), technology consultant and a product owner. Instead of having one scientific partner or consortium, our product owner interacts with the Dutch scientific community and is very active in attracting external users. In principle, our portfolio is well suited for Scrum, but the project budget and the team are too small to implement Scrum fully. We use Scrum artifacts like product backlog, and Scrum events such as daily stand-ups with RSEs, weekly meetings (with the Product Owner), and monthly retrospectives. Right now, our team is mainly concerned with a solid architecture of the product, and it might need several iterations until the team can start working on actual development. We are using GitHub Project (beta) as our Kanban board across the projects to put together a product backlog. We are hoping to get to development cycle soon and are open to adopt other Scrum artifacts like sprint backlog and increment (even pushing features as fast as possible) in future.'' Portfolio discipline: cross-disciplinary.

\subsection{Collectives}
\subsubsection{Collective:} ``Our collective was originally composed of four members who were working on very similar projects. These projects were focused on the same scientific domain and were using similar technologies that facilitated exchange between the members of the collective. The group grew to ten RSEs and is now down to six members. This fluctuation in size does not impact the functioning of the collective as most members work independently on their own project. However, keeping cohesion with a smaller group has been more difficult than when the group was larger. The organization of the collective is mostly done through our morning stand up where we exchange about our current work and help each other out when possible. The stand-up also serves as social moment that is especially important when working from home. We sometimes organize one-week sprints where the entire collective works together on the same project. The sprints are organized via a simple Kanban board, one member or an external collaborator take the role of product owner and one team member usually serves as Scrum master.'' Portfolio discipline: Material sciences.

\sidebox{\mylen}{lightgray}{white}{RSE's Lingo}{
  
\textbf{Agile}: A general term for a set of frameworks and methods based on principles described in Manifesto for Agile Software Developments \cite{agile-manifesto}.

\textbf{Scrum}: A framework for developing products (e.g., software) in a complex environment focusing on productivity and (incremental) delivery of the products.

\textbf{Stand-up}: A short time-limited meeting for each team members to share a brief status update on their work and issues impeding the work progress.

\textbf{Retrospective}: A time-limited meeting in which team discusses the work process in the team and identifies improvements for the team work and members individually.

\textbf{Sprint}: A predefined time period during which a team collaboratively works on a given objective.

\textbf{Pair programming}: an Agile software development technique, in which two engineers write and review a code together.

\textbf{Kanban board}: a visual board that depicts the work to be done and at which stage various tasks are. This is a very useful tool for lightweight team work management.

\textbf{Project (or sprint) backlog}: a prioritized list containing all the tasks the team plans to work during the course of the project (or the sprint). Various frameworks provide their own rules on who maintains it (also when and how).
}

\subsection{Agile Teams}
\subsubsection{Agile Team 1:} ``Our team consists of seven members (five engineers, a Scrum master and a project manager) and focuses mainly on big data and health projects, so both domain and technology are overlapping between projects. Each team member has one or two projects in which they act as lead engineer. Mostly, the team members work individually on their ‘own’ project, but there is ‘just-enough’ collaboration with the rest of the team. To make this work we have several meetings, that are inspired by several agile frameworks, but heavily adapted to our way of working: 1) A short planning meeting at the start of our two-week sprint in which we indicate on a high level what everyone is going to work on in their ‘own’ projects, how much time they will spend on activities outside of the team, and how much time they have left to help out on other projects within the team. 2) A review meeting at the end of the sprint in which each team member gives an overview of the progress that they made. 3) A retrospective meeting, right after the review meeting. 4) A stand-up meeting every morning. Occasionally we organise sprints where we focus on progressing in a single project with two or three team members.'' Portfolio discipline(s): Life Sciences, Health and Computer Science.

\subsubsection{Agile Team 2:} ``Our Team consists of six members, though the composition has changed quite a bit over the last few years. We based our initial process on Scrum but have changed our practice so much over time that it can no longer be regarded as Scrum by far. One of the team members organizes team meetings, but we try to rotate chairing meetings and taking notes among all team members. Other than that, we have no roles in the team. We do several projects, where each project is led by one of the team members. Almost all team members lead one or more projects. We try to work on all projects with all team members, although some members may contribute more to certain projects than others. Our team does quite a few projects that all require regular work and are hard to plan, so we struggled with the sprint model as defined in the Scrum framework. As an alternative, we switched to Kanban, where all tasks flow from left to right on our board and there are no sprints. At any time, there are items from multiple projects on the board. We try to ensure there is enough work on the board but limit the things that are in progress at any time to make sure we finish items. The project lead for a project adds items for that project to the backlog, and the team discusses what to pull onto the board. In addition, we have a retrospective every two weeks to discuss personal and team matters, and a daily stand-up to organize our work and keep everyone informed. After the daily stand-up we have a teamwork timeslot, used for either a review (demo) session within the team or for analysing backlog items. In the analysing session, we define what, how, who and acceptance criteria of an item, something like refinement or grooming. The difference with Scrum framework is that these sessions are unplanned. We do it whenever it is needed. Also, we plan some review and planning sessions with stakeholders separately.'' Portfolio discipline: Environment and Sustainability.

\subsection{Scrum Teams}
\subsubsection{Scrum Team 1:} ``Our team started as a project team but expanded and evolved into a pure Scrum team when the original project finished. Before the pandemic, we worked in the same office room. At the start of pandemic lockdown, we were five RSEs in the team including a certified Scrum master. We tried to adhere to the Scrum framework, some elements worked well, and others were simply too intense. For example, showing sprint board at the stand-ups did not work very well in terms of granularity. Pair programming worked well, also in remote settings; it is both chaotic and motivating. We worked in sprints, and only on one project per sprint. For smaller projects, the team split up into two sub-teams due to budget limitations. We made sure that the master branch is in working condition and followed best practices (working with pull requests and peer-reviewed code). At the end of some sprints, we created a project demo, and combined it with the sprint review. Scrum worked very well for projects with defined deliverables such as EU H2020 project. The product owner had a clear idea of what needs to be done and set clear priorities. In exploratory projects, it is hard to adhere to Scrum because project goals were more open ended, and it affected the teamwork. We noted a discrepancy between what eScience projects need and what Scrum framework can offer. For the exploration phase, it is best to work with one or two RSEs (including product owner), since it takes a bit of time to work out concise planning including deliverable and best suited tools or technologies. Working with the entire team on one project in a sprint meant going through project budget very quickly. This also resulted in a gap of a month to six weeks in between sprints on the same project, and we needed to put more efforts to get familiar with the project again.'' Portfolio discipline: Environment and Sustainability, Life Sciences, Computer Science.

\subsubsection{Scrum Team 2:} ``Our team is somewhere between Scrum and project team types, because it follows Scrum but is formed to work on one specific project. Due to the small budget, each RSE in our team can only work one day a week on this project. That is why the team members are also members of other teams in the Center, who in turn are involved in the other projects. Although roles in the team are ambiguous, we have a product owner who also does exploratory scientific work and handles user stories related work, and one team member acts as a Scrum master. All members have roughly the same expertise, thus, anyone can potentially work on any item. Almost all elements of Scrum are adopted and work well: the team works in sprints, starting with planning sessions, ending with review and retrospective meetings. A sprint duration is usually three weeks. Instead of daily stand-ups, our team has a stand-up twice a week. Despite the small budget, it is important to have stand-ups often; otherwise, a time gap between a stand-up and a moment when a team member can work on the project, tends to be too large to contribute fruitfully to the development process. Our team works towards minimal viable product (MVP) and our product maintains a Gantt chart (with priorities and deadlines), which is then reflected in the product and sprint backlogs. This chart is used for planning sessions to keep track of the team overall progress on the project. We define one goal/theme per sprint, and then everyone is working roughly on the same product feature. The work is focused on the sprint goal, which makes it more interesting for the team members to share their progress during stand-ups. This way of working is better for our team spirit and contributes to the feeling of mutual accomplishment as a team by the end of a sprint. Our review meetings are about what the team has achieved, and the retrospectives are about the work process. We make sure that any code or document is completed at the end of any sprint, and everything is peer reviewed by a team member. There was just one official release so far, but other sprints produced something that could have also been released. However, one third of the budget was devoted to the exploratory work (reading, researching), and these results are difficult to release immediately.'' Portfolio discipline: Computer Science.

\section{What Works When for Whom?}
Our experience has confirmed that there is no universal format that works in every situation. Each format has its own characteristics, and suitability depends on project needs and personality traits of a project team.

\subsection{Teams and Projects}
Due to the wide range of funding instruments we use, our projects vary in scope, duration, and budget (in terms of hours). Small explorative projects seem to fit better in unstructured teams such as Project Teams or Collectives. This format allows for ample time to clearly identify what the end goal of the project should be and explore various strategies and approaches. These explorative tasks are better performed by one or two engineers as they require a significant burn in period. These unstructured formats, however, do not facilitate the development of final products and often focus on prototyping and/or specific improvement of existing software.

Large projects naturally fit better in structured teams such as Agile or Scrum teams. The large contributions necessary for these projects require the high degree of synchronization that a tight team dynamic can provide. Team members are continuously aware of each other's activities allowing them to quickly alleviate bottlenecks and to ensure that their respective contributions are aligned. Smaller projects with a high degree of similarity can also benefit from being clustered together within such a team. When the overlap between these projects is sufficient, team members can easily contribute to several projects emulating a large project with smaller components. In addition, projects that are more product oriented also fit very well in structured teams. There, the tight collaboration between team members improves code design and maintainability. These structured teams generally require more planning and therefore an additional project management effort.

\subsection{Teams and People}
Another aspect that should be carefully considered, is that different formats fit better with the personal preferences of different individuals: a format that is inspiring and productive for some, may be completely unnatural and counterproductive for others. Attention to personal preferences, continuous reflection on the working format, and flexibility are some of the key ingredients for structuring a good team.

Unstructured teams such as Project Teams offer a lot of independence to the team members. Each member can plan their work independently and decide in which direction the project should go. This offers a lot of freedom to the research engineers that can rapidly develop prototypes and quickly explore and test ideas. However, these unstructured teams provide little cohesion between the team members that may feel isolated in their work. In addition, members of unstructured teams find little support in their teammates as their involvement in each other's work is limited.

Structured teams provide a very cohesive working environment and give a feeling of belonging to the team members. This allows team members to truly support each other and to distribute responsibilities among all the team members. However, the structure of the team limits the freedom and independence of each team member as they are fully committed to the work of the team. This can in the end decrease the sense of ownership of the team members.

\section{Making the best of both worlds}
All team structures briefly presented above have strengths and weaknesses and can provide either the perfect working environment or a never-ending hell for different people. Teams with a more rigid way of working, e.g., employing Scrum, are typically better equipped to build a finished product. However, they also require more management that comes at a cost. Agile teams have design-develop cycles which typically iterate every two weeks – for exploratory work, a quicker iteration cycle is beneficial~\cite{boehm86}. Unstructured teams are, by their nature, more flexible and can have shorter iteration cycles and therefore are more suitable for exploratory work.

All projects must deal with time, quality and money (budget) constraints~\cite{atkinson99} and make choices to find the right balance for that project: a more flexible team structure might require less budget to operate but may take longer to produce a polished product; a more structured team will burn through the budget faster but can produce higher quality product once development goals are sufficiently clear. Regardless of team structure, it is crucial to have a team member who will liaise with project partners and can set clear priorities for the deliverables.

One of the main lessons learned from experimenting with different teams is that there is no structure that is clearly superior to the others, each structure has its advantages and disadvantages. Our own understanding of what works for each team continues to evolve as we try different ways of working and each time are faced with different challenges. This is partially due to the research nature of scientific projects we work on.

Frequent and open discussions between members as well as with their line manager are crucial to identify personal preferences and find the best team for everyone. We should therefore let each team find that for themselves instead of unifying the inner working of every team. To ensure teams have the right skill set to self-organize we have developed a course on ‘Teamwork for Research Software Development’ that we will teach internally~\cite{teamwork-lesson}.

Having these diverse types of teams working alongside has even been a great asset for the eScience Center. As an example, the Integrated Omics project~\cite{omics} started as a very research oriented academic project for the exploration of machine learning techniques for understanding the interactions between microbes and human cells. A Project Team constituted of a single engineer was set up to work on the project. During the exploration phase the Project Team experimented with word2vec~\cite{mikolov13} a method originating from natural language processing. The success of the prototype prompted a consolidation effort of the initial code that was carried out by a Scrum team. This Scrum team significantly improved the code design and maintainability of the initial prototype enabling the adoption of the methods by a large community of bioinformaticians~\cite{spec2vec,matchms}. This success would have been impossible without the collaboration between structured and unstructured teams within the same organization.

\section{ACKNOWLEDGMENT}

It would have been impossible to create and gather all the information without the work of the employees of the Netherlands eScience Center. Many people have experimented with diverse types of teams and have been involved in the discussion regarding the pros and cons of each model. The authors would therefore like to thank all their colleagues for the many contributions they had and for making the Netherlands eScience Center such a great place to work at.

\bibliographystyle{unsrturl}
\bibliography{CsMag_template}

\end{document}